\documentclass{article}
\usepackage{latexsym,graphicx}
\newcommand{\be}{\begin{equation}}
\newcommand{\ee}{\end{equation}}
\usepackage[left=3.0cm,top=2.5cm,right=2.0cm,bottom=2.0cm]{geometry}
\usepackage{textcomp}
\usepackage{amsmath}
\usepackage{amssymb}
\usepackage{graphicx}
\begin{document}
\begin{center}
\large{\bf{Two Fluid Scenario in Bianchi Type-I Universe}} \\
\vspace{5mm}
\normalsize{ G. K. Goswami$^1$, Meena Mishra$^2$, Anil Kumar Yadav$^3$, Anirudh Pradhan$^4$ }\\
\vspace{2mm}
\normalsize{$^1$ Department of Mathematics, Kalyan P. G. College, Bhilai - 490 006, C. G., India}\\
\vspace{2mm}
\normalsize{$^2$ Department of Mathematics, Swami Shri Swaroopanand Saraswati Mahavidyalya,\\
Hudco, Bhilai - 490 006,C.G., India}  \\
 \vspace{2mm}
 \normalsize{$^2$ Department of Physics, United College of Engineering \& Research,\\
Greater Noida - 201310, India}  \\
 \vspace{2mm}
\normalsize{$^3$ Department of Mathematics, Institute of Applied Sciences and\\
Humanities, G L A University,  Mathura - 281 406, Uttar Pradesh, India } \\
 \vspace{2mm}
\normalsize{$^1$ Email: gk.goswam9@gmail.com} \\
\vspace{2mm}
\normalsize{$^2$ Email: minamishra18@gmail.com} \\
 \vspace{2mm}
 \normalsize{$^3$ Email: abanilyadav@yahoo.co.in} \\
 \vspace{2mm}
 \normalsize{$^4$ E-mail: pradhan.anirudh@gmail.com} \\
\end{center}
\vspace{3mm}
\begin{abstract}
  In this paper, we study a Bianchi  type -I  model of universe  filled with barotropic and
  dark energy(DE) type fluids. The present values of cosmological parameters such as Hubble constant $H_0$,
barotropic, DE and anisotropy energy parameters  $(\Omega_{m})_0$,
$(\Omega_{de})_0$ and $(\Omega_{\sigma})_0 $ and Equation of State(EoS)
parameter for DE  ($\omega_{de}$) are statistically estimated  in two ways by
taking 38 point data set of Hubble parameter H(z) and 581 point data set of
distance modulus of supernovae in the range $0\leq z \leq 1.414$. It is found that
the results agree with the  Planck result [P.A.R. Ade, et al., Astron. Astrophys. 594 A14 (2016)] and more latest result
obtained by Amirhashchi and Amirhashchi [H. Amirhashchi and S. Amirhashchi, arXiv:1811.05400v4 (2019)]. Various physical properties
such as age of the universe, deceleration parameter etc have also been
investigated.
\end{abstract}


\section{Introduction;}
 SN Ia observations \cite{ref1}$-$\cite{ref4} confirm the fact that our
observable universe is accelerating at present. This surprising discovery is a
break through in the field of observational cosmology and had lead to a
 presence of an unknown dark energy(DE) fluid that opposes gravitational attraction.
It is a common perception that DE has  positive energy density and negative
pressure so that it creates acceleration in the universe. Although, it violate the
strong energy condition (SEC), yet  provides an elegant description of transition
of universe from deceleration  to cosmic acceleration (Caldwell et al.\cite{ref5}). In the framework of general relativity, the dynamics of dark energy could be understand through it's equation of state parameter which is defined as $\omega^{(de)} = p^{(de)}/\rho^{(de)}$ where $p^{(de)}$ and $\rho^{(de)}$ are the pressure and energy density of dark energy component respectively. It is well known that $\omega^{(de)} = -1$ represents the standard $\Lambda$CDM model of universe. \\

After CMB experiment, It has been now confirmed that the matter distribution inside the present universe
is on whole isotropic but early universe had not such smooth picture i.e. it was anisotropic near the singularity point. So, one has to assume anisotropy in the background of evolving process of current universe. Off late, spatially homogeneous and anisotropic cosmology had been a matter of
interest to the cosmologists. Recently, Akarsu et al. \cite{ref6} have constructed Bianchi
type I model (BT-I) as natural extension of  the standard $\Lambda$CDM model.
Amirhashchi  and  Amirhashchi \cite{ref7}  have investigated three DE models
for  flat and curved FLRW and BT- I space times and put constraints on
cosmological parameters using Gaussian processes and  MCMC method.  In
other papers \cite{ref8,ref9}, they developed BT-I Universe with Type Ia Supernova
and H(z) Data and have probed DE in the scope of BT-I space time.  Mishra et al.
\cite{ref10} investigated the role of anisotropic components on the DE and the
dynamics of the universe in Bianchi-V string cosmological model. In another
papers \cite{ref11,ref12}, they have also discussed Bulk viscous embedded BT- I
dark energy models. Recently Rashid et al. \cite{ref13} have also developed
anisotropic DE model.  More information and
references regarding BT-I DE models can be found in  Goswami et al. \cite{ref14}$-$\cite{ref20}. Some  important applications of BT-I cosmological models in the framework of general relativity and modified theories of gravitation are given in Refs. \cite{Kumar/2007,Singh/2008,Singh/2008a,Akarsu/2010,Yadav/2019bjp,Sharma/2018,Yadav/2019mpla}. \\

 In this paper, we study a BT -I  model of universe  filled with barotropic and
  DE perfect fluids. The present values of cosmological parameters
  such as Hubble constant $H_0$, barotropic, DE and anisotropy energy parameters
  $(\Omega_{m})_0$, $(\Omega_{de})_0$ and $(\Omega_{\sigma})_0 $ and
  Equation of State(EoS) parameter for DE $\omega_{de}$ are statistically estimated
 in two ways by taking 38 point data set of Hubble parameter H(z) and 581 point data set of
distance modulus of supernovas in the range $0\leq z \leq 1.414$. It is found that
the results agrees with the  Planck findings \cite{ref21} and more latest results
due to  Amirhashchi and  Amirhashchi \cite{ref22}. The contents of the paper in brief are as follows : In section 2, we have described
the field equation for BT- I universe. In section 3 and 5,  Hubble and energy
parameters were estimated in the two ways on the basis of 38 data set of H(z)
and a distance modulus data set of 581 Supernovas. In section 4, luminosity
distance, distance modulus and apparent magnitude in our model have been
formulated. In section 6, Various physical properties such as age of the universe,
deceleration parameter etc have also been investigated. Finally the concluding remarks are presented in section 7.

\section{ Field equations for Bianchi Type I Universe}
We consider  a general BT- I metric
\begin{equation}
\label{eq1}
ds^2 = dt^2-a^2dx^2-b^2dy^2-c^2dz^2,
\end{equation}
where a, b and c are scale factors along spatial directions and it depend on time only.\\

Let the universe be filled with  two
type of fluids: one is barotropic and other creating dark energy. We assume
that suffix  m stands for matter and  de for dark energy.\\
The energy momentum
tensor(EMT) has two components i.e. $  T_{ij} = T_{ij}(m)+T_{ij}(de) $. The
followings are the EMTs of the contents of the universe, $T_{ij}(m)=\left(\rho_m +
p_m\right)u_{i}u_{j}-p_m g_{ij}$ and
$T_{ij}(de)=\left(\rho_{de}+p_{de}\right)u_{i}u_{j}-p_{de} g_{ij}$. For co-moving
co-ordinates
 $u^{\alpha}=0; \; \; \alpha=1,2~\&~ 3$, where $ g_{ij}u^iu^j=1$.
 The Einstein field equations are
 \begin{equation}
 \label{eq2}
 R_{ij}-\frac{1}{2}Rg_{ij}= - 8\pi GT_{ij},
 \end{equation}
  where we have taken velocity of light as unity. The field equation (\ref{eq2}) in terms of line element (\ref{eq1}) and EMTs
  described above are solved as follows.[See \cite{ref14} for details]\\
   $a^2 = bc$, $b = ad$, $c = ad^{-1}$,
 $2\frac{a_{44}}{a}+\frac{a^{2}_{4}}{a^2}+ (\frac{d_4}{d})^2 = -8\pi G(p_{de}+p_m)$,
  $\frac{a^{2}_4}{a^2}- (\frac{d_4}{d})^2= 8\pi G(\rho_m+\rho_{de})$ and
  $(\frac{d_4}{d})_4+ 3\frac{a_4d_4}{ad}=0$, which on integration gives
  $\frac{d_4}{d} =\frac{k}{a^3}$, where k is an arbitrary constant of integration.
  The Hubble's parameter H in this model is as follows
 $H = \frac{1}{3}(\frac{a_4}{a}+\frac{b_4}{b}+\frac{c_4}{c})= \frac{a_4}{a}$.
 Finally we get following field equations for BT-I anisotropic universe

\begin{equation}
\label{eq3}
2\frac{a_{44}}{a}+\frac{a^{2}_{4}}{a^2} = -8\pi G(p_{de}+p_m+p_{\sigma})
\end{equation}
\begin{equation}
\label{eq4}
H^2 = \frac{a^{2}_4}{a^2} = \frac{8\pi G}{3}(\rho_m+\rho_{de}+\rho_{\sigma})
\end{equation}
where we have considered anisotropy terms $\frac{d_4}{d}$ appearing in the field
equations as anisotropy energy represented by suffix $\sigma$ whose pressure
and density are given by
\begin{equation}
\label{eq5}
p_{\sigma}=  \rho_{\sigma}=\frac{k^{2}}{8\pi Ga^{6}}.
  \end{equation}
 During  course of its evolution, the universe had gone through a stage where matter density and pressure were equal.
  That stage is called stiff matter filled universe. So we can say that anisotropy energy  behaves like stiff matter.
The energy conservation law for our model is as follows
$T_j^{ij}=\dot{\rho}+3H(p+\rho)=0$, where
 $\rho=\rho_{m}+\rho_{de}+\rho_{\sigma}$   and $p=p_{m}+p_{de}+p_{\sigma}$.
 We see that $\dot{\rho_{\sigma}}+3H(p_{\sigma}+\rho_{\sigma})=0$  holds separately. So we have
$\frac{d}{dt}{(\rho_{m}+\rho_{de})}+3H(p_{m}+p_{de}+\rho_{m}+\rho_{de})=0.$ We
assume that dark energies does not interact with barotropic matter , so that they
are conserved simultaneously i.e. $(\rho_{m})_4+3H(p_{m}+\rho_{m})=0$ and
$(\rho_{de})_4+3H(p_{de}+\rho_{de})=0$. The equations of states are as follows
$p_{m}=\omega_{m}\rho_{m}$, where $\omega_{m}$ are constants. For matter in
form of radiation $\omega_{m}=\frac{1}{3},\rho_{m} \varpropto a^{-4}.$ Present
universe is dust filled for which $\omega_{m}=0,
 \rho_{m}\varpropto a^{-3}.$ Now we use the following relation between scale factor a and red shift z,  $\frac{a_0}{a} =
 1+z$. This gives $ \rho_{\sigma}= (\rho_{\sigma})_0  [\frac{a_0}{a}]^6= (\rho_{\sigma})_0
(1+z)^6.$  Similarly   $\rho_{de} = (\rho_{de})_0
  [\frac{a_0}{a}]^{3(1+\omega_{de})}=(\rho_{de})_0 (1+z)^{3(1+\omega_{de})},$
   where $\omega_{de}$ is equation of state parameter for dark energy which is considered as constant for present epoch.
  We take $p_m=0$ for dust filled universe and define following energy parameters~$ \Omega_{m}=\frac{\rho_{m}}{\rho_{c}},~~
 \Omega_{de}=\frac{\rho_{de}}{\rho_{c}},~~\&~~
 \Omega_{\sigma}=\frac{\rho_{\sigma}}{\rho_{c}}$, where   $\rho_{c}=\frac{3H^{2}}{8\pi G}$.\\

  The field equations (\ref{eq3}) and (\ref{eq4})
  now take following form
 \begin{equation}\label{eq6}
 (2q-1) H^2 = 3 H^2_0 \left(\omega_{de}(\Omega_{de})_0(1+z)^{3(1+\omega_{de})} + (\Omega_{\sigma})_0 (1+z)^6    \right)
 \end{equation}
 and
 \begin{equation}
 \label{eq7}
  H^2 =  H^2_0 \left( (\Omega_{m})_0(1+z)^{3} + (\Omega_{de})_0(1+z)^{3(1+\omega_{de})} + (\Omega_{\sigma})_0 (1+z)^6 \right)
 \end{equation}
 where $q = \frac{\ddot{a}}{aH^2}$ is deceleration parameter(DP).\\

 The  relationship amongst the energy parameters are obtained from Eq.(\ref{eq7}) as
  $$(\Omega_{m})_0+(\Omega_{de})_0+(\Omega_{\sigma})_0=1$$

\section { Hubble and energy parameters  based on 38 data set of H(z) }

 So many astrophysical scientists  \cite{ref24} $-$  \cite{ref28} estimated Hubble constant $H_0$,
   as  $ 72 \pm 8 ,~~  69.7^{+4.9}_{-5.0},~~ 71 \pm 2.5
 ,~~ 70.4^{+1.3}_{-1.4},~~ 73.8 \pm 2.4 $ and
  $~~ 67 \pm 3.2 $ in the unit $ km s^{-1} Mpc^{-1} $  respectively, with the help of  Hubble Space Telescope (HST) ,Cepheid variable
   observations,  gravitational lensing , WMAP seven-year data and WMAP results with Gaussian priors ,  infrared camera
  and galactic cluster  data's  respectively. One may refers to Kumar \cite{ref29}, Sharma et al.
  \cite{Sharma/2019} and Yadav et al. \cite{Yadav/2019} for detail. We consider a observed data set of 38  Hubble
   parameter Hob(i) in Gyr$^{-1}$ unit  with standard deviations $\sigma(i)$  for
   different red shifts. These were imported from  Farook at el  \cite{ref23}.
    For corresponding theoretical value of H(z),  we use Eq.(\ref{eq7}), in which $H_0, (\Omega_{m})_0,
   (\Omega_{\sigma})_0$ and $\omega_{de}$ are unknown. It is desired to estimate values of these parameters statistically
   by getting Chi square given by
   \begin{equation}
   \label{eq8}
     \chi^{2}[H_0, (\Omega_{m})_0,\omega_{de}] = \sum _{i=1}^{i=38}[(Hth(i) - Hob(i))^2/\sigma(i)^2],
    \end{equation}
   where Hth (i)'s are theoretical values of Hubble parameter as per Eq.(\ref{eq7}) and $\sigma(i)$'s are errors in the observed values of H(z).\\

  We take ($0.066 \leq H_0 \leq  0.076$) , ($0.1 \leq (\Omega_{m})_0 \leq 0.5$)
   and EoS parameter $\omega_{de}$ in the range ( -1.3,~ -0.8).
   As at present anisotropy is very mere, we take $ (\Omega_{\sigma})_0$ = 0.0002.
   It is found that $\chi^{2}$= 33.22 i.e. 87.43 \% is minimum  for $H_0 = 0.068~ Gyr^{-1}$ = 66.6 Km/sec/Mpc , $(\Omega_{m})_0$ =0.26 ,  $\omega_{de}$ = -0.83.
  and $(\Omega_{de})_0$ = 0.7398. Now we present a error bar graph as figure 1 which have 38 observational Hubble data (OHD) points (left panel) and H(z) + BAO (right panel) with possible errors as bars and a curve representing corresponding theoretical value of H(z) given by Eq(\ref{eq7}). We also note that the solid black line represent the best fit curve of derived model and dashed blue and green lines represent the corresponding $\Lambda$CDM model ($\omega^{(de)} = -1$) respectively.\\

We have taken $ H_0, (\Omega_{m})_0$, $( \Omega_{\sigma})_0 $ and $(\Omega_{de})_0$ as estimated statistically on the basis of minimum  $\chi^{2}$.
     Figure 2 represents 1$\sigma$, 2$\sigma $ Confidence regions in the ($\Omega_{m}, \omega_{de} $) plane. Inside these regions red ellipse shows our estimated values.
     These figures  show that observed and theoretical values are close to each other.
     \begin{figure}[ht]
     \centering
    \includegraphics[width=7cm,height=6cm,angle=0]{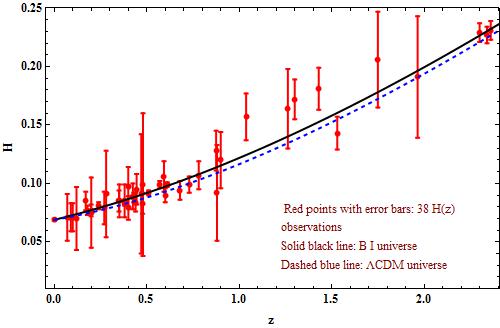}
    \includegraphics[width=7cm,height=6cm,angle=0]{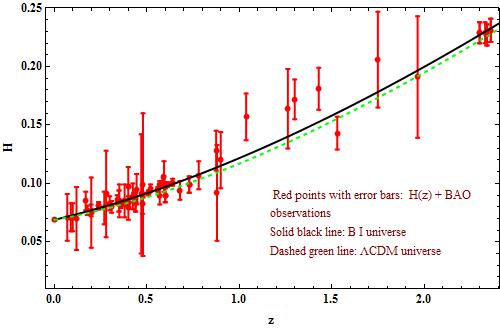} \\
    \caption{Hubble paprmeter $-$ red shift error bar plot with 38 OHD points (left panel) and with H(z) + BAO (right panel) for $H_0 = 0.068~ Gyr^{-1}$ = 66.6 km/s/Mpc, $(\Omega_{m})_0$ =0.26, $\omega_{de}$ = -0.83 and $(\Omega_{de})_0$ = 0.7398.}
      \end{figure}
       \begin{figure}[ht]
      \centering
      \includegraphics[width=7cm,height=6cm,angle=0]{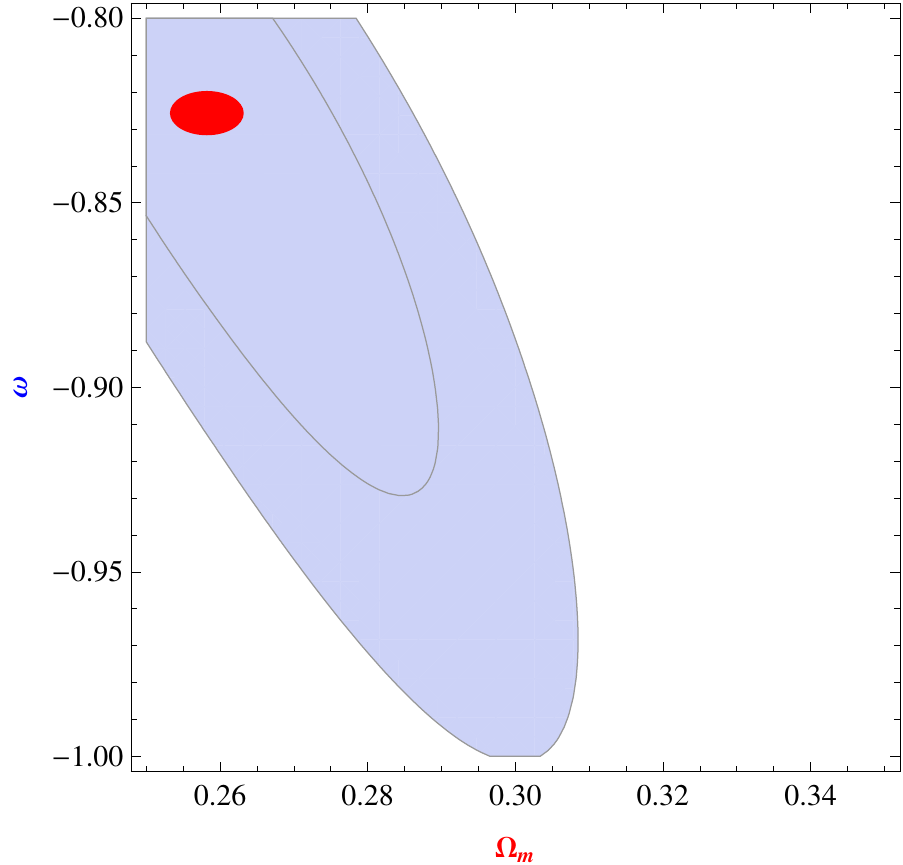} \\
      \caption{The likelihood contours at 1$\sigma$, 2$\sigma $ Confidence regions in $\Omega_{m} - \omega$ plane by bounding our model with 38 OHD points. Here $\omega_{de}$ is constant and it is taken as $\omega_{de} = \omega$.}
      \end{figure}
   \newpage
\section{  Luminosity Distance, Distance modulus and Apparent magnitude in our model }
   In our earlier work  \cite{ref14,ref15} for B-I universe,
   Luminosity distance $D_L$, Distance modulus $\mu$ and Apparent magnitude $m_b$ of any distant luminous object are
   obtained as $ D_{L}=\frac{c(1+z)}{H_{0}}\int^z_0\frac{dz}{h(z)}$, $\mu = M-m_{b}= 5log_{10}\left(\frac{D_L}{Mpc}\right)+25$
   and $ m_{b}=16.08+ 5log_{10}\left[\frac{1+z}{.026} \int^z_0\frac{dz}{h(z)}\right]$ where $h(z)= \frac{H(z)}{H_0}.$
   There fore in the present model, these  physical quantities are given as
 \begin{equation}
 \label{eq9}
   D_{L}=\frac{(1+z)}{H_{0}}     \int^z_0\frac{dz} {\sqrt{(\Omega_m)_0(1+z)^3+(\Omega_{de})_0 (1+z)^{3(1+\omega_{de})}+(\Omega_{\sigma})_0 (1+z)^{6}}},
  \end{equation}

 \begin{equation}\label{eq10}
   \mu = 25+ 5log_{10}\left(\frac{(1+z)}{H_{0}}\int^z_0\frac{dz}{\sqrt{(\Omega_m)_0(1+z)^3+(\Omega_{de})_0 (1+z)^{3(1+\omega_{de})}+(\Omega_{\sigma})_0 (1+z)^{6}}} \right)
  \end{equation}
and
 \begin{equation}\label{eq11}
 m_b = 16.08+ 5log_{10}(\frac{1+z}{.026} \int^z_0\frac{dz}{\sqrt{(\Omega_m)_0(1+z)^3+(\Omega_{de})_0 (1+z)^{3(1+\omega_{de})} +(\Omega_{\sigma})_0 (1+z)^{6}}}  )
 \end{equation}
\section{Hubble and energy parameters based on  a distance modulus  data set of 581 Supernovas}
 We consider a observed data set of   distance modulus of 581 Supernovas   with standard deviations ~$\sigma  SN1a(i)$  for
   different red shifts in the range $z\leq 1.414$. These were imported from Pantheon compilation \cite{ref30}.
    For corresponding theoretical value of  $\mu_ {th}$,  we use Eq.(\ref{eq10}), in which $H_0, (\Omega_{m})_0,
   (\Omega_{\sigma})_0$ and $\omega_{de}$ are unknown. It is desired to estimate values of these parameters statistically
   by getting Chi square given by
\begin{equation}\label{eq12}
\chi^{2}((\Omega_{m})_0,(\Omega_{\sigma})_0 ,\omega_{de})= \sum _{i=1}^{Length SN1aData} \frac{\mu_{th} ((\Omega_{m})_0,( \Omega_{\sigma})_0 ,\omega_{de})(i)-\mu_ {obs}(i))^2}{\sigma  SN1a(i)^2}
\end{equation}
 We take ($0.066 \leq H_0 \leq  0.076$) , ($0.1 \leq (\Omega_{m})_0 \leq 0.5$)
   and EoS parameter ($-1.3\leq \omega_{de}\leq -0.8$).
   Like as before, we take $ (\Omega_{\sigma})_0$ = 0.0002.
   It is found that $\chi^{2}$= 562.227 i.e. 96.7$\%$ is minimum for  $H_0$ = 70.0097,   $(\Omega_m)_0 $ = 0.279 and
$\omega_{de}$ = -1.00654.
 Now we present a error bar graph as figure 3 which has 581
data points with possible errors as bars and a curve representing corresponding
theoretical value of $\mu(z)$ given by Eq(\ref{eq10}). In figure 3, the solid black line represents the best fit curve of the model under consideration while dashed blue line corresponds to the $\Lambda$CDM model ($\omega^{(de)}= -1$). We have taken   $ H_0,
(\Omega_{m})_0$, $( \Omega_{\sigma})_0 $ and $(\Omega_{de})_0$ as
estimated statistically on the basis of minimum  $\chi^{2}$. Figure(4) represents
1$\sigma$, 2$\sigma $ Confidence regions in the ($\Omega_{m}, \omega_{de} $)
plane. Inside these regions red ellipse shows our estimated values. These figures
show that observed and theoretical values are close to each other.
\begin{figure}[ht]
    \centering
    \includegraphics[width=7cm,height=6cm,angle=0]{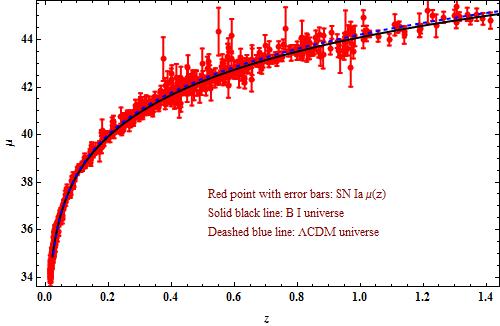} \\
    \caption{ Distance modulus ($\mu$) $-$ red shift error bar plot for $H_0$ = 70.0097, $(\Omega_m)_0 $ = 0.279, $(\Omega_{de})_0 $ = 0.7208  and $\omega_{de}$ = -1.00654.}
\end{figure}
\begin{figure}[ht]
    \centering
    \includegraphics[width=7cm,height=6cm,angle=0]{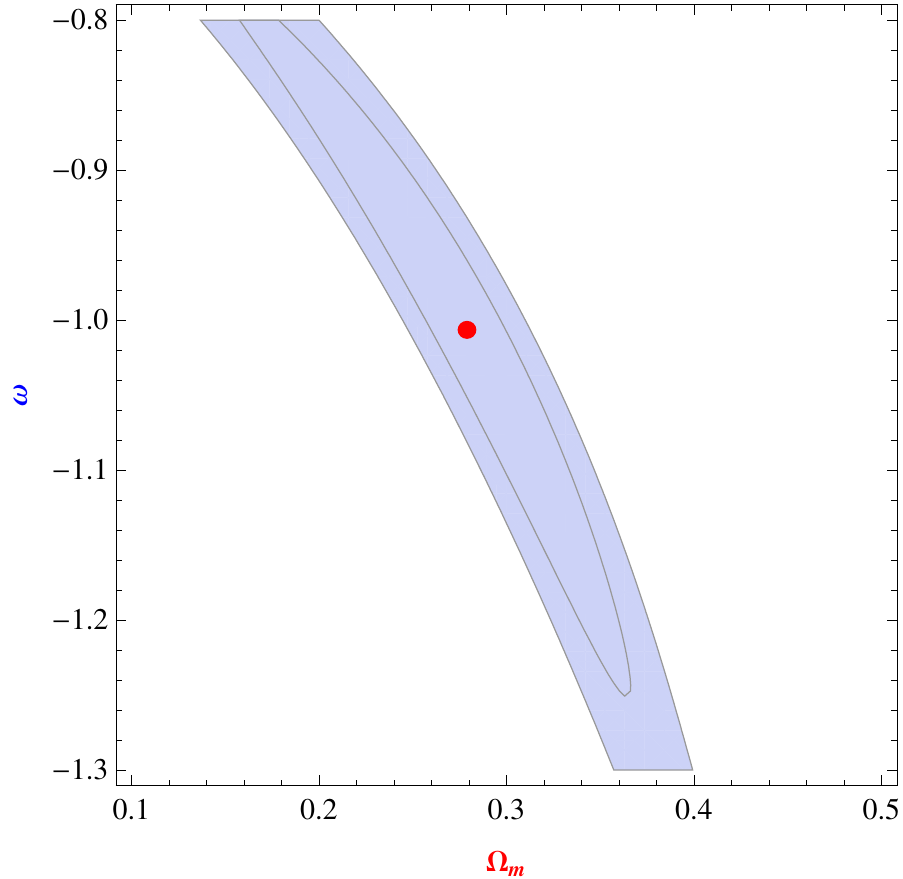} \\
    \caption{1$\sigma$, 2$\sigma $ Confidence regions and our estimated point inside the red ellipse in $\Omega_{m} - \omega$ plane by bounding our model with 581 SN Ia data. Here $\omega_{de}$ is constant and it is taken as $\omega_{de} = \omega$.}
\end{figure}

\section{Present  age of the universe}
We obtained the present age of universe as follows\\
$$t=\intop_{0}^{t}dt=\intop_{0}^{a}\frac{da}{aH}=\intop_{0}^{z}\frac{dz}{(1+z)H}$$\\
This implies that
\begin{equation}\label{eq13}
 t_0 =\lim_{x\rightarrow\infty}{ \int_{0}^{x}\frac{dz}{ H_0(1+z)  \sqrt((\Omega_m)_0(1+z)^3+(\Omega_{de})_0 (1+z)^{3(1+\omega_{de})}+(\Omega_{\sigma})_0 (1+z)^{6})}}.
\end{equation}
We see that $ t_0 \rightarrow 0.95296 H_0^{-1} $ for high  red shifts of order
$10^5$, where we have taken  $H_0$ = 70.0097,   $(\Omega_m)_0 $ = 0.279
and $\omega_{de}$ = -1.00654. Now $ H_0^{-1}=13.9976~ Gyrs$, so the present
age of universe comes to $ t_0= 13.339~Gyrs$ for our model. If we calculate
$t_0$ on the basis of our results $H_0 = 0.068~ Gyr^{-1}$ = 66.6 Km/sec/Mpc ,
$(\Omega_{m})_0$ =0.26 ,  $\omega_{de}$ = -0.83.
  and $(\Omega_{de})_0$ = 0.7398 as per 38 pt.Hubble parameter estimation, we
  get age of the universe as $t_0=13.7711~Gyrs$. The empirical value of present age of the universe is
  $t_{0}=13.73_{-.17}^{+.13}$  as per WMAP data\cite{ref31}. Thus  present age of universe obtained by us is
very close to observed one especially with respect to 38 OHD. Hubble parameter
estimation.  The figure 5  describes variation of time with red shift.
 \begin{figure}[ht]
 \centering
   \includegraphics[width=8cm,height=6cm,angle=0]{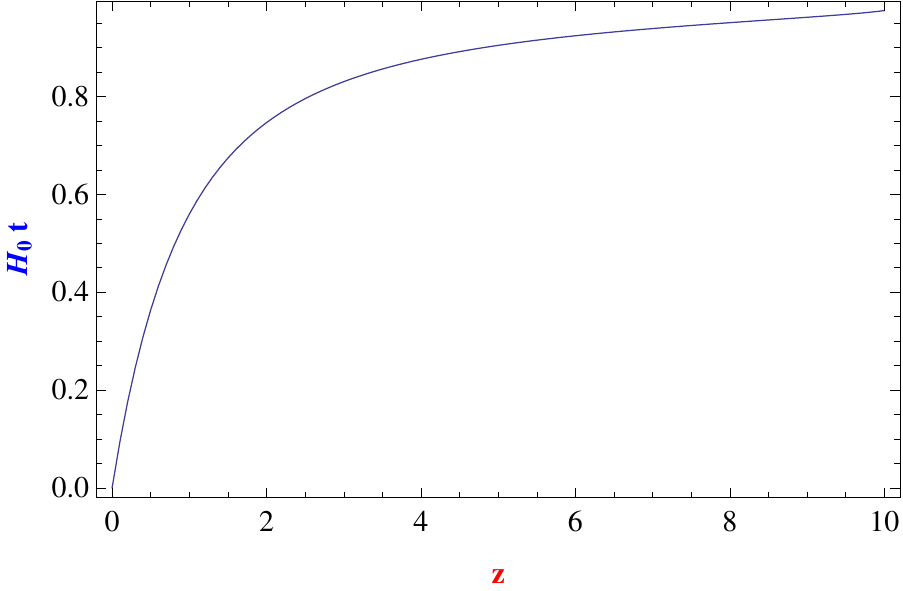}
    \caption{Red shift z versus time t plot for $H_0$ = 70.0097,   $(\Omega_m)_0 $ = 0.279
and $\omega_{de}$ = -1.00654.}
\end{figure}
\subsection{ Deceleration Parameter}
The deceleration parameter(DP) is obtained from Eqs.(\ref{eq6}) and (\ref{eq7}) as
 \begin{equation}\label{eq14}
  2q = 1 + 3\frac{\omega_{de} (\Omega_{de})_0 (1+z)^{3(1+\omega_{de})}-\frac{1}{3}(\Omega_{\sigma})_0 (1+z)^6}{\sqrt{(\Omega_{m})_0 (1+z)^3+(\Omega_{\sigma})_0  (1+z)^6+ (\Omega_{de})_0 (1+z)^{3(1+\omega_{de})}}}.
  \end{equation}
 It's present value is given as~  $2(q)_0=1+3\omega_{de} (\Omega_{de})_0-(\Omega_{\sigma})_0$.\\

In absence of dark energy, our model represent an decelerating universe.  Dark energy has negative pressure ($\omega_{de}<0$), so it makes the universe accelerating. For  $H_0$ = 70.0097, $(\Omega_m)_0 $ = 0.279 and $\omega_{de}$ =
   -1.00654, the present value of DP is obtained as $(q)_0 =  -0.58837$. The present value of DP on the basis of our results $H_0 = 0.068~ Gyr^-1$ ,
   $(\Omega_{m})_0$ =0.26 ,  $\omega_{de}$ = -0.83 and $(\Omega_{de})_0$
   =0.7398 as per 38 points. Hubble parameter estimation comes out to be equal to
   $(q)_0$= -0.421335. The following figure(6) shows  how deceleration parameter q varies over red shift z. It is interesting to see that there are two transition red shift in this model $z_t= 1.448$
 and $z_t=4.18$. This means that our universe had gone two times though the accelerating phase. Duration of  present phase is $0 \leq z \leq 1.448$ and in the past it was
 $z \geq 4.18$. This shows that structure formation era is $1.448 \leq z \leq 4.18$ and inflation might have taken place at $z \geq 4.18$. Figure(6)
     describes the whole evolution of the universe. It covers the main three phases of the universe, Inflation,
    structure formation and the present accelerating phase.

 \begin{figure}[ht]
  \centering
    \includegraphics[width=8cm,height=6cm,angle=0]{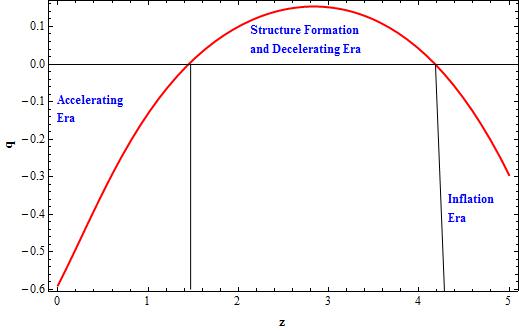}
    \caption{Deceleration parameter q  over red shift z for $H_0$ = 70.0097,   $(\Omega_m)_0 $ = 0.279 and $\omega_{de}$ = -1.00654.}
\end{figure}
\subsection{Particle Horizon}
Let us consider a light ray form a source along x-direction. Proper distance of the source will be
$a_0 x$. Let we are receiving light signal at certain time $t_0$. It might have
transmitted in the past at certain time say $t_p$ from the source, then proper
distance of the source form us will be given
 by $ r = a_0 \int_{t_p}^{t_0}\frac{dt}{a(t)}$.\\
 \begin{figure}[ht]
\centering
   \includegraphics[width=8cm,height=6cm,angle=0]{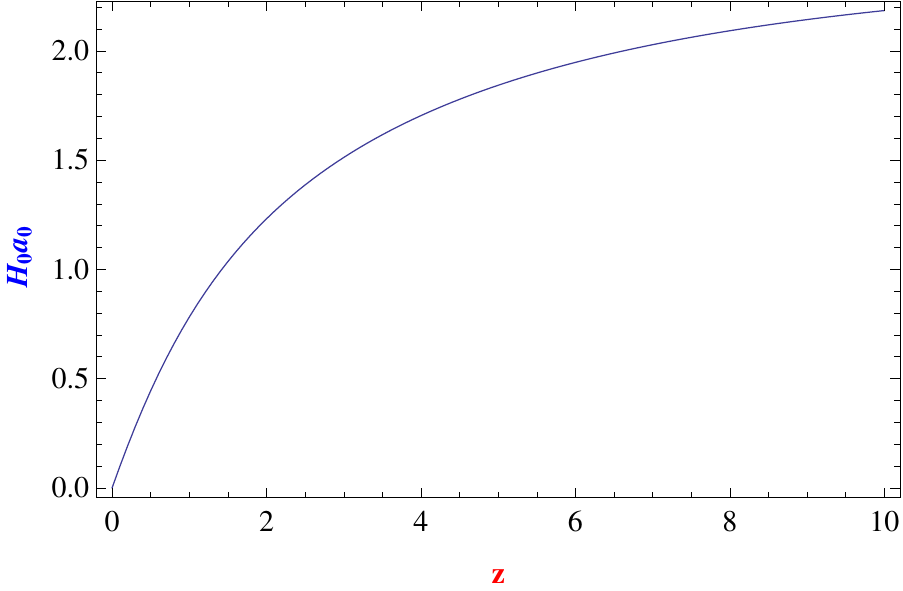}
    \caption{Red shift z versus proper distance plot for $H_0$ = 70.0097,   $(\Omega_m)_0 $ = 0.279
and $\omega_{de}$ = -1.00654.}
\end{figure}

The Particle Horizon $R_P$  is defined as the $ \lim_{t_p \rightarrow 0} a_0{ \int_{t_p}^{t_0} \frac{dt}{a(t)}} =\lim_{z \rightarrow \infty }
\int_{0}^{z}\frac{dz}{H(z)}$

 $\therefore$
\begin{equation}\label{eq29}
 R_P =\lim_{x\rightarrow\infty}{ \int_{0}^{x}\frac{dz}{ H_0 \sqrt((\Omega_m)_0(1+z)^3+(\Omega_{de})_0 (1+z)^{3(1+\omega_{de})}+(\Omega_{\sigma})_0 (1+z)^{6})}}.
\end{equation}
We see that $ R_P \rightarrow 2.43419 H_0^{-1} $ for high  red shift of order
$10^5$, where we have taken  $H_0$ = 70.0097,   $(\Omega_m)_0 $ = 0.279
and $\omega_{de}$ = -1.00654.  The value of $R_P$  on the basis of our results
$H_0 = 0.068~ Gyr^{-1}$ ,
   $(\Omega_{m})_0$ =0.26 ,  $\omega_{de}$ = -0.83 and $(\Omega_{de})_0$
   =0.7398 as per 38 OHD. Hubble parameter estimation comes out to be equal to
   $R_P \rightarrow 2.48215 H_0^{-1} $.  For FLRW model it is given as $
R_{P}\sim\frac{2}{H_0}$.\\

 The figure 7  describes variation of proper distance with red shift.

\section{Conclusion}
In this paper, we have investigated the two fluid scenario in Bianchi type I space-time. It is worth to mention that in the literature, two fluid interacting dark energy models in BT-I space-time are available \cite{Kumar/2011,Yadav/2012,Yadav/2016}. But the mechanism for solving of field equations in present model is altogether different from the mechanism used in Refs. \cite{Kumar/2011,Yadav/2012,Yadav/2016}. We also estimate the present values of cosmological parameters of derived model by using 38 OHD points and 581 SN Ia data. We summaries our finding with the help of the following table. We have also displayed
observational  data's due to Planck for the purpose of comparison. Figure 6 of the our work is very interesting. It describes the whole evolution of the universe from it's beginning to present epoch. It covers the main three phases of the universe, Inflation, structure formation and the current accelerating phase.

    \begin{center}
        \begin{tabular}{|c|c|c|c|c|c|}
            \hline
            Cosmological   &  Values as  &   Values  as & Planck \\
             Parameters    &   per 38 OHD  &  per 581 SN Ia& results \\
              at present     &   & &           \\
             \hline
            $(\Omega_{de})_0$   & 0.7398 & 0.7208 & 0.6911 \\
            $(\Omega_m)_0$      & 0.26 & 0.279&0.3089 \\
             $(\Omega_\sigma)_0$  & 0.0002& 0.0002& 0 \\
            $\omega_{de}$   &  -0.83 & -1.00654 &-1.019\\
            $H_0$              &  66.6 & 70.0097& 67.74    \\
            $Age~ t_{0}$  &13.7711 & 13.339~Gyrs  & 13.799\\
             $R_{P}$ & 2.48215$H_0^{-1}$ & 2.43419 $H_0^{-1}$  &--- \\
              $(q)_0$ &    -0.421335          & -0.58837   &---   \\
              \hline
        \end{tabular}
    \end{center}
    We also quote the latest results due to   Amirhashchi and  Amirhashchi \cite{ref22}
    $H_0 = 69.9 \pm 1.7, (\Omega_m)_0 = 0.279^{+0.014}_{ -0.016}, (\Omega_{de})_0   = 0.721^{+0:016}_{-0.014}$ and $ z_t = 0.707\pm 0.034$.
 \section*{Acknowledgement}
The authors (G. K. Goswami  \&  A. Pradhan) sincerely acknowledge the
Inter-University Centre for Astronomy and Astrophysics (IUCAA), Pune, India for
providing facilities where part of this work was completed during a visit.

     \end{document}